\documentclass[twocolumn]{aa}

\usepackage{graphicx,natbib,txfonts,verbatim}
\bibpunct{(}{)}{;}{a}{}{,}

\begin{document}

\title{Turbulent cross-field transport of non-thermal electrons in coronal
loops: theory and observations}

\author{N. H. Bian, E. P. Kontar, and A.L. MacKinnon}

\offprints{N.H. Bian \email{nicolas.bian@glasgow.ac.uk}}

\institute{Department of Physics \& Astronomy, University of Glasgow, G12 8QQ, United Kingdom}

\date{Received ; Accepted }

\abstract{A fundamental problem in astrophysics is the interaction between
magnetic turbulence and charged particles. It is now possible to
use \emph{Ramaty High Energy Solar Spectroscopic Imager (RHESSI)} observations
of hard X-rays (HXR) emitted by electrons to identify the presence of turbulence
and to estimate the magnitude
of the magnetic field line diffusion coefficient at least in dense coronal flaring loops.}
{We discuss the various possible regimes of cross-field transport of non-thermal electrons resulting
from broadband magnetic turbulence in coronal loops. The importance of the Kubo number $K$
as a governing parameter is emphasized and results applicable in both the large and small Kubo number limits are collected.}
{Generic models, based on concepts and insights developed in the statistical
theory of transport, are applied to the coronal loops and to the interpretation
of hard X-ray imaging data in solar flares. The role of trapping effects, which become
important in the non-linear regime
of transport, is taken into account in the interpretation of the data.}
{For this flaring solar loop,
we constrain the ranges of parallel and perpendicular correlation lengths
of turbulent magnetic fields and possible Kubo numbers. We show that a substantial amount of magnetic fluctuations with
energy $\sim 1\%$ (or more) of the background field can be inferred
from the measurements of the magnetic diffusion
coefficient inside thick-target coronal loops.}
{}


\titlerunning{cross-field transport}

\authorrunning{Bian et al}

\maketitle

\section{Introduction}
Solar flares provide many observational challenges
for crucial aspects of high-energy astrophysics, including
energy release, particle acceleration and transport
in magnetized plasmas. In the standard flare scenario, magnetic energy stored in the
corona is released via plasma heating, bulk motions and particle acceleration.
Thanks to hard X-ray (HXR) imaging spacecraft
such as \emph{Yohkoh}/HXT \citep{1991SoPh..136...17K}
and \emph{RHESSI} \citep{2002SoPh..210....3L}, high-resolution spatial and spectral
diagnostics of energetic particles \citep{1999Ap&SS.264..129S,2002SSRv..101....1A, 2003AdSpR..32.1001L,
2005AdSpR..35.1675B,2006SSRv..124..233L}
have proven to be vital for our understanding of the physics
of the solar corona.

Turbulence, an important element of the solar flare scenario, is
believed to be associated with various physical processes,
from the triggering of fast magnetic reconnection to particle acceleration and transport.
Many particle acceleration models rely on the presence of electromagnetic fluctuations
during flares and it has been shown that stochastic
acceleration can effectively energize and accelerate a large number of electrons
and ions \citep{1987SoPh..113..195M,
1992ApJ...398..350H, 1994ApJS...90..623M, 2009ApJ...692L..45B, 2010ApJ...712L.131P,bian}.
The precise origin of the turbulence in the acceleration region
of flaring loops is still unclear but it has been suggested that it
could be associated with current sheets \citep{1987ApJ...317..900C, 1997ApJ...485..859S,
2006A&A...452.1069L} and/or with reconnection outflows \citep{1994ApJ...425..856L}.
Independently of its origin, if turbulence has a significant impact
on particle acceleration, it must
also be expected to manifest itself via the transport of particles, including HXR-emitting electrons.
This opens the route of using HXR data to characterize turbulent processes involved
in particle acceleration and transport during solar flares.

A step forward in this direction was recently taken by \citet{2011ApJ...730L..22K},
who developed a method for determining
the magnetic diffusion coefficient in flaring loops. Their approach
was inspired by a study of \citet{2008ApJ...673..576X}, which analyzes the variation
of the HXR source size along the guiding field of the loop
as a function of energy, see also \citet{2009ApJ...706..917P}.
Specifically, \citet{2011ApJ...730L..22K} have shown that the size of the HXR source in the direction
perpendicular to the magnetic field is also a growing function of
energy. These observations strongly suggest cross-field mobility of non-thermal electrons
inside flaring loops.

Owing to their high speed and small Larmor radius, the cross-field
transport of energetic electrons in loops is likely to be dominated
by perpendicular magnetic fluctuations and by the resulting
wandering of the magnetic field lines. The perpendicular transport
of magnetic field lines is usually quantified in terms of a diffusion coefficient
$D_{m}$ and estimates of this diffusion coefficient were obtained in \citet{2011ApJ...730L..22K} through
imaging observations of HXR-emitting electrons.

The magnetic diffusion coefficient $D_{m}$ depends on three quantities, which are the relative level of the turbulent magnetic fluctuations
and their parallel and perpendicular correlation lengths: $B_{\perp}/B_{0}$, $\lambda_{z}$, $\lambda_{\perp}$, respectively.
Because the dimension of the magnetic diffusion coefficient is a length, dimensional analysis
gives that $D_{m}\approx (\lambda^{2}_{\perp}/\lambda_{z})K^{\gamma}$. The number $K$ is the Kubo number
defined as $K=(B_{\perp}/B_{0})(\lambda_{z}/\lambda_{\perp})$ \citep{1998PhRvE..58.7359V, 2000EJPh...21..279B, 2000PPCF...42....1B, 2000PhyA..280...99Z}. It characterizes the magnetic turbulence.
Its importance stems from the fact that it is the only non-dimensional parameter
that enters the equation describing the perpendicular transport of magnetic field lines.
When the Kubo number is small, i.e. $K \ll 1$, the turbulent transport of field lines is well-described by
the quasilinear approximation,
which predicts that $D_{m}\approx (\lambda^{2}_{\perp}/\lambda_{z})K^{\gamma}$ with $\gamma=2$.
This result is identical to the case when the turbulence is slab, i.e. $\lambda_{\perp}=\infty$. In other words, when $K\ll1$, which is the domain
of applicability of the quasilinear approximation, the magnetic diffusion coefficient is independent of $\lambda _{\perp}$
and scales as the second power of the relative level of fluctuations $B_{\perp}/B_{0}$, i.e. $D_{m}\approx \lambda_{z}(B_{\perp}/B_{0})^{2}$.
From the HXR measurements of $D_{m}$, \citet{2011ApJ...730L..22K} gave constraints on the level of turbulent
magnetic fluctuations for a specific event, assuming that the magnetic turbulence is slab or equivalently $K\ll 1$.
 However, it is established that anisotropy of turbulence,
with $\lambda_{\perp}\ll \lambda_{z}$, is prevalent
in magnetized plasmas. Moreover, when the Kubo number is large, i.e. $K\gg 1$, the quasilinear theory
fails, the magnetic diffusion coefficient
no longer scales as $(B_{\perp}/B_{0})^{2}$ but instead $D_{m}\approx (\lambda^{2}_{\perp}/\lambda_{z})K^{\gamma}$ with $\gamma<1$.
Therefore, if we aim to relate measurements of the magnetic
diffusion coefficient to the relative level of magnetic fluctuations
produced by the turbulence inside flaring loops,
we need not only additional observational constraints on
the correlation lengths $\lambda_{z}$ and $\lambda_{\perp}$ but also
some theoretical predictions on the scaling of $D_{m}$ with $B_{\perp}/B_{0}$ when $K \gg 1$.

Here, we discuss the various regimes
of cross-field transport of non-thermal electrons resulting
from broadband magnetic turbulence in flaring coronal loops.
Results applicable in both the large and small $K$ limits are collected
and are applied to the interpretation
of hard X-ray imaging data.

\section{Perpendicular motion of energetic electrons
in coronal loops}
In the guiding-centre approximation, the perpendicular transport of
particles is described by
\begin{equation}
\frac{d\mathbf{r_{\perp}}}{dt}={\rm v}_{z}\frac{\mathbf{B}_{\perp}}{B_{0}}+\frac{1}{B_{0}}(\mathbf{E}_{\perp}\times \mathbf{b}_{0}),
\end{equation}
where the background magnetic field $\mathbf{B}_{0}$ is uniform and directed
along $\mathbf{z}(\equiv \mathbf{b}_{0})$,
$\mathbf{B}_{\perp}$ and $\mathbf{E}_{\perp}$ are magnetic and electric fluctuations
perpendicular to $\mathbf{B}_{0}$,
and ${\rm v}_{z}$ is the electron velocity parallel to the guiding field.
There are two contributions to the perpendicular transport of particle gyrocentres.
One contribution arises from electric field fluctuations,
which produce the $\mathbf{E}\times \mathbf{B}$-drift:
$\mathbf{\rm v}_{E}=(1/B_{0})\mathbf{E}_{\perp}\times \mathbf{b}_{0}$.
The other contribution comes from the magnetic field fluctuations which also produce a perpendicular drift given by
$\mathbf{\rm v}_{B}={\rm v}_{z}(\mathbf {B}_{\perp}/B_{0})$.

The effect of perpendicular electric field fluctuations is negligible
for the cross-field transport of non-thermal electrons in coronal
loops, which have ${\rm v}_{Te}\sim {\rm v}_{A}$, ${\rm v}_{Te}$ is the electron thermal speed
and ${\rm v}_{A}$ the Alfven speed. Indeed for magnetohydrodynamics (MHD) turbulence,
$E_{\perp}\sim {\rm v}_{A}B_{\perp}$
and therefore
${\rm v}_{B}/{\rm v}_{E}\sim {\rm v}_{z}/{\rm v}_{A}$.
The neglect of the $\mathbf{E}\times \mathbf{B}$ drift contribution to
perpendicular transport is thus justified provided ${\rm v}_{z}\gg {\rm v}_{A}$.
Indeed, \citet{2011ApJ...730L..22K} report  ${\rm v}_{A}\simeq 1000$~km/s
and ${\rm v}_{z} \simeq 50000$~km/s for electrons producing tens of keV X-rays.

Because the smallness of the $\mathbf{E}\times \mathbf{B}$ drift is verified for non-thermal
electrons in coronal loops, it means that the cross-field transport
is dominated by magnetic fluctuations.
As a consequence, the gyrocentre equation of motion
simplifies to
\begin{equation}
\frac{d\mathbf{r_{\perp}}}{dt}=v_{z}\frac{B_{\perp}}{B_{0}}.
\end{equation}
In other words, fast electrons in coronal loops tend to follow the field
lines because their Larmor radius is small (few centimeters for the coronal parameters) and because their
$\mathbf{E}\times \mathbf{B}$ drift is unimportant. Let us notice, however, that the electric contribution
to perpendicular transport becomes of the same order as the magnetic
contribution for thermal electrons.


\section{Magnetic field-line transport}

The above discussion shows that the cross-field transport of fast electrons
is dominated by the turbulent field-line wandering that is generated through
\begin{equation}
\frac{d\mathbf{r_{\perp}}}{dz}=\frac{\mathbf{B}_{\perp}}{B_{0}}.
\label{eq3}
\end{equation}
We
emphasize that Eq.(\ref{eq3}) has a Hamiltonian structure because $\mathbf{B}_{\perp}=\nabla A_{z}\times \mathbf{z}$,
where $A_{z}$ is the parallel component of the vector potential.

A control parameter of the problem is the Kubo number. This number appears
by writing the field-line equation (\ref{eq3}) in a form obtained after
normalizing $r_{\perp}$ and $z$ by $\lambda_{z}$ and $\lambda_{\perp}$, the parallel and perpendicular correlation
lengths of the magnetic perturbations. This form is
$d\mathbf{r_{\perp}}/dz=\mathbf{K}$ with $\mathbf{K}=(\lambda_{z}/\lambda_{\perp})(\mathbf{B_{\perp}}/{B_{0}})$.
Therefore the only non-dimensional parameter entering the equation describing the perpendicular transport of field line is
\begin{equation}
K=\frac{B_{\perp}}{B_{0}}\frac{\lambda_{z}}{\lambda_{\perp}},
\label{eq:kubo}
\end{equation}
which is called the Kubo number \citep{1998PhRvE..58.7359V, 2000EJPh...21..279B, 2000PPCF...42....1B, 2000PhyA..280...99Z}.

Without loss of generality, we focus on the dispersion of magnetic field lines in the $x$
direction given by
\begin{equation}
\frac{dx}{dz}=\frac{B_{x}}{B_{0}}.
\end{equation}
A similar equation for the $y$-displacement involves $B_{y}/B_{0}$ and it is assumed
that $B_{y}/B_{0}\sim B_{x}/B_{0}$.

The turbulent field is homogeneous with zero average,
$<B_{x}>=0$, where $<>$ denotes the ensemble average. The two-point \emph{Eulerian correlation function} of the
magnetic perturbation is given, it is
\begin{equation}
E(\mathbf{r_{\perp}},z)=<B_{x}(0,0)B_{x}(\mathbf{r}_{\perp},z)>.
\end{equation}
As an example, we may consider the following form
\begin{equation}\label{ex}
E(\mathbf{r_{\perp}},z,)=B^{2}_{x}\exp\left(-r^{2}_{\perp}/\lambda^{2}_{\perp}\right)\exp(-z/\lambda_{z}),
\end{equation}
which depends only on two arguments, $r_{\perp}=\mid \mathbf{r_{\perp}}\mid$ and $z$, because of the homogeneity and isotropy
in the perpendicular plane.

A main goal of the theory is to determine the variation with $z$ of $<(\Delta x)^{2}>$, $\Delta x$
being the field-line displacement.
To this purpose, it is convenient to introduce a running diffusion coefficient,
which is defined as $D_{m}(z)=d<(\Delta x)^{2}>/2dz$.
An important property of this running diffusion coefficient is that it is related to the \emph{Lagrangian correlation
function} by the Taylor formula
\begin{equation}\label{run}
D_{m}(z)=\frac{1}{B_{0}^{2}}\int_{0}^{z}L(z')dz'.
\label{eq:taylor}
\end{equation}
Here, the notation
\begin{equation}
L(z)\equiv <B_{x}(0)B_{x}(z)>,
\label{eq:lagrangian}
\end{equation}
is used for the Lagrangian correlation, where $B_{x}(z)\equiv B_{x}(\mathbf{r}_{\perp}(z),z)$
is obtained through $\mathbf{r}_{\perp}(z)$ by integration of the field-line
equations.
When the integral in Eq.(\ref{run}) converges to a non-zero constant in the limit $z\rightarrow \infty$,
i.e. $D_{m}(z\rightarrow \infty)\rightarrow D_{m}$, it follows from the definition of the running
diffusion coefficient that
\begin{equation}
<(\Delta x)^{2}>=2 D_{m}z.
\end{equation}
This expression does not determine the magnetic diffusion coefficient $D_{m}$ but simply states that the field-line displacement
follows a standard diffusive process.
Deviations from the standard diffusive transport can occur
whether $D_{m}(z\rightarrow \infty)\rightarrow 0$ or $D_{m}(z\rightarrow \infty)\rightarrow \infty$.
For instance, when $<(\Delta x)^{2}>\propto z^{\alpha}$ with $0<\alpha<1$, the field-line transport
is said to be sub-diffusive, while for $<(\Delta x)^{2}>\propto z^{\alpha}$ with $\alpha>1$,
the transport is super-diffusive.

Although it is the Eulerian correlation $E(\mathbf{r_{\perp}},z,)$ or the spectrum of magnetic
fluctuations that is assumed to be a known function, this is instead the Lagrangian correlation
function $L(z)$
that determines the running diffusion coefficient, and hence, also the mean square displacement of the field lines through the
Taylor formula (\ref{eq:taylor}).
Consequently, the whole difficulty of the turbulent transport theory resides in the determination
of the Lagrangian correlation function corresponding to a given Eulerian correlation function, for instance of the form
(\ref{ex}). A widespread procedure that relates the Lagrangian correlation to the Eulerian one is the Corrsin approximation
(see Section 3.2). The Corrsin approximation is equivalent to the quasilinear approximation when the Kubo number is small.
However, this procedure fails to account accurately for the role of the non-linearity in the field line equation, and in particular trapping effects, which become important
when the Kubo number is large.

\subsection{Slab turbulence}
The determination of the magnetic diffusion coefficient is greatly simplified when the magnetic turbulence is slab, i.e. when the magnetic perturbations
are a function of $z$ only, because the Eulerian and Lagrangian correlation
functions coincide in this case. Therefore
the magnetic diffusion coefficient reads
\begin{equation}
D_{m}=\frac{1}{B^{2}_{0}}\int_{0}^{\infty}dz'<B_{x}(0)B_{x}(z')>\simeq \lambda_{z}\left(\frac{B_{x}}{B_{0}}\right)^{2},
\end{equation}
and this magnetic diffusion coefficient exists provided $\lambda_{z}$ is finite.
This is also the well-known expression for the quasilinear diffusion coefficient \citep{1966ApJ...146..480J, 1978PhRvL..40...38R}.

\subsection{The Corrsin approximation}

In general, $B_{x}(\mathbf{r}_{\perp},z)$ is a function of $\mathbf{r_{\perp}}$ that makes
the field line equation non-linear. As a result it is a difficult task to express the Lagrangian correlation function $L(z)$
(Equation \ref{eq:lagrangian}), which enters the Taylor formula (\ref{eq:taylor})
in terms of the known Eulerian correlation function $E(x,y,z)$.
By definition, $L(z)=\int d\mathbf{r}_{\perp}<B_{x}(0,0)B_{x}(\mathbf{r}_{\perp},z)\delta [\mathbf{r}_{\perp}-\mathbf{r}_{\perp}(z)]>$
and $E(\mathbf{r_{\perp}},z)=<B_{x}(0,0)B_{x}(\mathbf{r}_{\perp},z)>$.
The vast majority of turbulent transport theories are based on the assumption that the propagator $
\delta [\mathbf{r}_{\perp}-\mathbf{r}_{\perp}(z)]$ is equal to its average over the statistics of the fluctuations,
i.e. $\delta [\mathbf{r}_{\perp}-\mathbf{r}_{\perp}(z)]=<\delta [\mathbf{r}_{\perp}-\mathbf{r}_{\perp}(z)]>$.
This independence hypothesis goes back to \citet{1959AdGeo...6..441C} and allows
the Lagrangian correlation function to be written as
 \begin{equation}
L(z)=\int_{-\infty}^{+\infty} d\mathbf{r}_{\perp} E(\mathbf{r}_{\perp},z)P(\mathbf{r}_{\perp},z),
\end{equation}
where $P(\mathbf{r}_{\perp},z)\equiv <\delta [\mathbf{r}_{\perp}-\mathbf{r}_{\perp}(z)]>$ is the probability
for a field-line to make a perpendicular excursion
from $0$ to $\mathbf{r}_{\perp}$  over a distance $z$. In the Corrsin approximation
the Lagrangian correlation is obtained as
a weighted average of the Eulerian correlation that involves the probability
distribution function $P(\mathbf{r}_{\perp},z)$.

When the Kubo number is small, i.e. $K\ll 1$, the right-hand side of the field line equation (\ref{eq3})
is also small. Hence, it is possible to make the approximation that $P(r_{\perp},z)\sim \delta(r_{\perp})$.
Consequently, it follows from (12) that the Lagrangian correlation is given by the Eulerian correlation
around $r_{\perp}=0$:
 \begin{equation}
 L(z)\sim E(0,z).
 \end{equation}
This is equivalent to the quasilinear approximation \citep{1966ApJ...146..480J, 1978PhRvL..40...38R}, which
yields the following expression for the magnetic field-line diffusion coefficient:
 \begin{equation}
 D_{m}\simeq \lambda_{z} \left(\frac{B_{x}}{B_{0}}\right)^{2}=\frac{\lambda_{\perp}^{2}}{\lambda_{z}}K^{2}.
 \label{eq:D_quasilinear}
 \end{equation}
This is just the restatement of Eq.(11). This scaling of the magnetic diffusion coefficient as the second power of the Kubo number
is generally valid provided the Kubo number is much smaller than unity, $K\ll 1$, including
the case of slab turbulence. The quasilinear diffusion coefficient scales as the second power
of the relative level of magnetic fluctuations.

The substitution
$\lambda_{z}=\lambda_{\perp}^{2}/D_{m}$ in Equation (\ref{eq:D_quasilinear})
provides the following expression for the magnetic diffusion coefficient:
 \begin{equation}
D_{m}\simeq \lambda_{\perp}\left(\frac{B_{x}}{B_{0}}\right)=\frac{\lambda_{\perp}^{2}}{\lambda_{z}}K,
 \end{equation}
a relation which was originally proposed by \citet{1979ppcf....1..649K}. These scaling of the diffusion coefficient as the first
and second power of the Kubo number
(first and second power of the relative level of magnetic fluctuations) can both be obtained under the Corrsin independence hypothesis when the probability
distribution function $P(x,y,z)$ satisfies the diffusion equation
\begin{equation}
\frac{\partial P}{\partial z}=D_{m}\left(\frac{\partial^{2}P}{\partial x^{2}}+\frac{\partial^{2}P}{\partial y^{2}}\right),
\end{equation}
with the condition that $P(r_{\perp},0)=\delta(r_{\perp})$.
Indeed, the substitution of the Gaussian solution for $P(x,y,z)$ or $P(r_{\perp},z)$, which depends on $D_{m}$, into
\begin{equation}
D_{m}=\frac{1}{B_{0}^{2}}\int_{0}^{\infty}dz\int_{-\infty}^{+\infty} dxdy E(x,y,z)P(x,y,z),
\label{eq:D_2D}
\end{equation}
provides an integral equation for $D_{m}$. Alternatively, (\ref{eq:D_2D}) can be written in terms
of the spectral energy density of the magnetic fluctuations as
\begin{equation}\label{cor2}
D_{m}=\frac{1}{B_{0}^{2}}\int d^{3}\mathbf{k}\mid B_{x}\mid^{2}_{\mathbf{k}}\frac{D_{m}k_{\perp}^{2}}{(D_{m}k_{\perp}^{2})^{2}+k_{z}^{2}},
\end{equation}
with $<B_{x}(\mathbf{k})B_{x}(\mathbf{k}')>=\mid B_{x}\mid^{2}_{\mathbf{k}}\delta(\mathbf{k}+\mathbf{k'})$.
A characteristic result of this kind of analysis is
an implicit relation for $D_{m}$ rather than an explicit expression.
This procedure was named "renormalization" in the review article by \citet{1993PhyU...36.1020B},
see also \citet{2009ApJ...692L..45B}.
The asymptotic limits $K\ll 1$ and $K\gg1$ of Eq.(\ref{eq:D_2D}) or Eq.(\ref{cor2}),
recover the scaling of the diffusion coefficient as second and first power of the Kubo
number respectively. Indeed, when $D_{m}k^{2}_{\perp}\gg k_{z}$, i.e. $K\gg 1$, the Corrsin approximation gives
$D^{2}_{m}=(1/B_{0}^{2})\int d^3\mathbf{k}\mid B_{x}\mid ^{2}_{\mathbf{k}}/k_{\perp}^{2}$.
Here, an essential result
is that the magnetic diffusion coefficient remains finite and is given
by $D_{m}\simeq \lambda_{\perp} (B_{x}/B_{0})$, even
 for a strictly two-dimensional turbulence. An expression that interpolates the $K\ll 1$ and $K\gg1$
regimes of transport obtained \emph{under the Corrsin approximation} can be written as
\begin{equation}
D_{m}=\frac{\lambda_{\perp}^{2}}{\lambda_{z}}K^{2}(1+K)^{-1},
\label{dminterp}
\end{equation}
which gives $D_{m}=(\lambda_{\perp}^{2}/\lambda_{z})K^{2}$ for $K<<1$ and $D_{m}=(\lambda_{\perp}^{2}/\lambda_{z})K$ for $K>>1$.

\subsection{Magnetic field-line trapping}

 A major problem with the Corrsin approximation is precisely that it predicts a non-zero
diffusion coefficient, independent of $\lambda_{z}$, for 2D turbulence when $K=\infty$.
Indeed, a 2D turbulence with $\lambda_{z}=\infty$
is incapable of producing a standard diffusion. The reason is that the field-line equation is fully integrable
and that the original Hamiltonian system for 2D perturbations cannot generate stochastic field lines.
Particles that follow the field lines and that are released on surfaces $A_{z}(x,y)={\rm const}$
that close on themselves (minima and maxima
of the flux function) will remain trapped inside flux tubes. In pure 2D turbulence, the majority
of field lines wind around flux tubes, which means that a) particles
that follow the field lines stay confined within the flux tubes, b) the mean-square
displacement of particles cannot grow with time so that the diffusion coefficient
has to be zero.

Nevertheless, even a weak parallel dependence of the turbulence on $z$ is able to produce
the opening of the closed contours, which releases field lines and
hence particle trajectories in the perpendicular plane; see Figure (1) in \citet{2010ApJ...711..997H}. A typical trajectory shows
an alternation of trapping and perpendicular displacement.
These perpendicular displacements occur when particles remain in regions of low absolute values
of the flux function, i.e. close to the magnetic separators, and the overall process is a diffusion.

Dimensionally, the scaling of $D_{m}$ with $K$, in the $K\gg1$ non-linear limit,
has to obey
\begin{equation}
D_{m}=\frac{\lambda_{\perp}^{2}}{\lambda_{z}}K^{\gamma}=(\delta B_{\perp}/B_{0})^{\gamma}\lambda_{z}^{\gamma-1}\lambda_{\perp}^{2-\gamma}
\end{equation}
with $\gamma<1$, in order for $D_{m}$ to vanish when $\lambda_{z}=\infty$.
The first estimate of $\gamma$ for $K\gg1$ was given by Gruzinov and co-workers based on an analogy with percolation
in a stochastic landscape \citep{RevModPhys.64.961}.
It yields $\gamma \simeq 2/3$; see also the discussion in \citet{2009PhRvE..79d6403M}.
This value appears to be valid for Eulerian correlation functions that decay
sufficiently fast. Balescu and co-workers have developed analytical methods yielding
important progress in the statistical theory
of transport \citep{1998PhRvE..58.7359V, 2000EJPh...21..279B, 2000PPCF...42....1B}.
An expression that interpolates the quasilinear and trapping regime
of transport can be written as
\begin{equation}
D_{m}=\frac{\lambda_{\perp}^{2}}{\lambda_{z}}K^{2}(1+K^{4/3})^{-1},
\label{dminterp}
\end{equation}
which gives $D_{m}=(\lambda_{\perp}^{2}/\lambda_{z})K^{2}$ for $K<<1$ and $D_{m}=(\lambda_{\perp}^{2}/\lambda_{z})K^{2/3}$ for $K>>1$.

The importance of anisotropy and trapping effects in the non-linear regime
of transport was recently discussed in the context of the propagation
of solar energetic electrons in the solar wind \citep{2010ApJ...711..997H}
and for the transport of thermal electrons in solar coronal loops \citep{2010ApJ...719.1912B}.
These considerations, which are here applied to the transport of non-thermal electrons in flaring loops,
 may lead to substantial variations
in the value of the turbulence level that is inferred by applying the turbulent
transport theory to the interpretation of data.

\section{Application to transport of fast HRX-emitting electrons in
thick-target coronal loops}

In a recent work we developed an approach for determining the magnetic diffusion coefficient
in thick-target loops. This is based on \emph{RHESSI} observations and the X-ray visibility
analysis \citep{2002SoPh..210...61H}. Once X-ray visibilities are fitted with Gaussian-curved ellipsoids,
the loop sizes clearly reveal \citep{2008ApJ...673..576X,2011ApJ...730L..22K}
that both the longitudinal (along the guide field) and latitudinal (across-the guide field)
extents of the HXR source, $L(\epsilon)$ and $W(\epsilon)$, are increasing functions of the photon energy $\epsilon$.
The energy-dependent loop-length is no surprise
within a thick-target scenario \citep{1971SoPh...18..489B}
because higher energy electrons can travel farther away from the region where they are
accelerated. As a result the HXR source appears to be longer at higher photon
energies \citep{2002SoPh..210..373B}, as is
often observed in the dense regions of the atmosphere \citep[e.g.][]{2002SoPh..210..383A,2006AdSpR..38..962M,2008A&A...489L..57K,2009ApJ...706..917P,2010ApJ...717..250K,2010ApJ...721.1933S}.

One important point here is the increase of the HXR source width with energy.
This indicates that transport of particles also occurs in the direction perpendicular to the mean magnetic field
of the loop. An other important point is that because of their high speed and small Larmor radius, the cross-field transport of
electrons is dominated by magnetic fluctuations
inside the loop.

While a fast electron emits in the HXR range, it also travels a distance given
by $z\simeq \epsilon^{2}/2K'n$ in the direction along the guide field,
$K'=2\pi e^{4}\ln \Lambda$, $\ln \Lambda\simeq 20$ is the Coulomb logarithm.
For an energy independent length $L_0$ of the acceleration
region, $L(\epsilon)$ is given by \citep{2008ApJ...673..576X} and can be well
approximated by
\begin{equation}
L(\epsilon)=L_{0}+\alpha_{z}\epsilon^{2},
\label{eq:l_eps}
\end{equation}
where $\alpha_{z}\simeq 1/(2K'n)$. As a result of the perpendicular transport of field lines,
the same electrons also make a perpendicular excursion given by $r_{\perp}\simeq\sqrt{2D_{m}z}$.
This produces the increase of $W(\epsilon)$ with energy. $W$ is the sum
of the acceleration region width $W_{0}$ and the part due to lateral transport:
\begin{equation}
W(\epsilon)=W_{0}+\alpha_{\perp}\epsilon,
\label{eq:w_eps}
\end{equation}
where $\alpha_{\perp}=\sqrt{2D_{m}\alpha_{z}}$.
By fitting Equations (\ref{eq:l_eps})-(\ref{eq:w_eps})
to the observed $L(\epsilon)$ and $W(\epsilon)$, it is possible to determine the values
of the parameters that enter these equations and, hence, to obtain the value
of the magnetic diffusion coefficient $D_{m}$, i.e. $D_{m}=\alpha_{\perp}^2/(2\alpha_{z})$.
\citet{2011ApJ...730L..22K} found $D_{m}\simeq 2\times 10^{7}$~cm,
and also $L_{0}\simeq 2\times 10^{9}$~cm and $W_{0} \simeq 5\times 10^{8}$~cm for the rising
phase of the flare.

From this measured value of $D_m$ we would ideally like to determine a value of $\delta B/B_{0}$ ($\delta B\equiv B_{\perp}$)
but this would require independent knowledge of
the correlation lengths $\lambda_z$ and $\lambda_\perp$. Nonetheless we are able to place interesting constraints on all
three parameters. First of all, we can take $\lambda_{z} < L_{0}$ and $\lambda_{\perp} < W_{0}$ because $L_{0}$
and $W_{0}$ are of the order of the integral scales of the visible loop. We must have $\delta B/B_0 < 1$ as well,
realizing that magnetic turbulence can power the whole flare for $\delta B\sim B_{0}$; see \citet{2011ApJ...730L..22K}.
Considering (\ref{dminterp}) with $\lambda_z$ held fixed,
we see that a lower limit to $\delta B/B_0$ is given by the quasilinear estimate,
\begin{displaymath}
\left(\frac{\delta B}{B_0}\right)^2 \, = \, \frac{D_m}{\lambda_z}
\end{displaymath}
and thus that $D_m < \lambda_z < L_0$. For a fixed value of $\lambda_z$ in this range, $\delta B/B_0$ is a decreasing function of $\lambda_\perp$
so the requirement $\delta B/B_0 < 1$ now sets a lower limit to $\lambda_\perp$, obtainable by rewriting Eq.~(\ref{dminterp}):
\begin{equation}
\lambda_\perp \, = \lambda_z \left( \frac{\delta B}{B_0} \right) \left( \frac{\lambda_z}{D_m}\left(\frac{\delta B}{B_0} \right)^2 - 1\right)^{-4/3}.
\label{eq:lperp}\end{equation}

\begin{figure*}
\centering
\includegraphics[width=\textwidth]{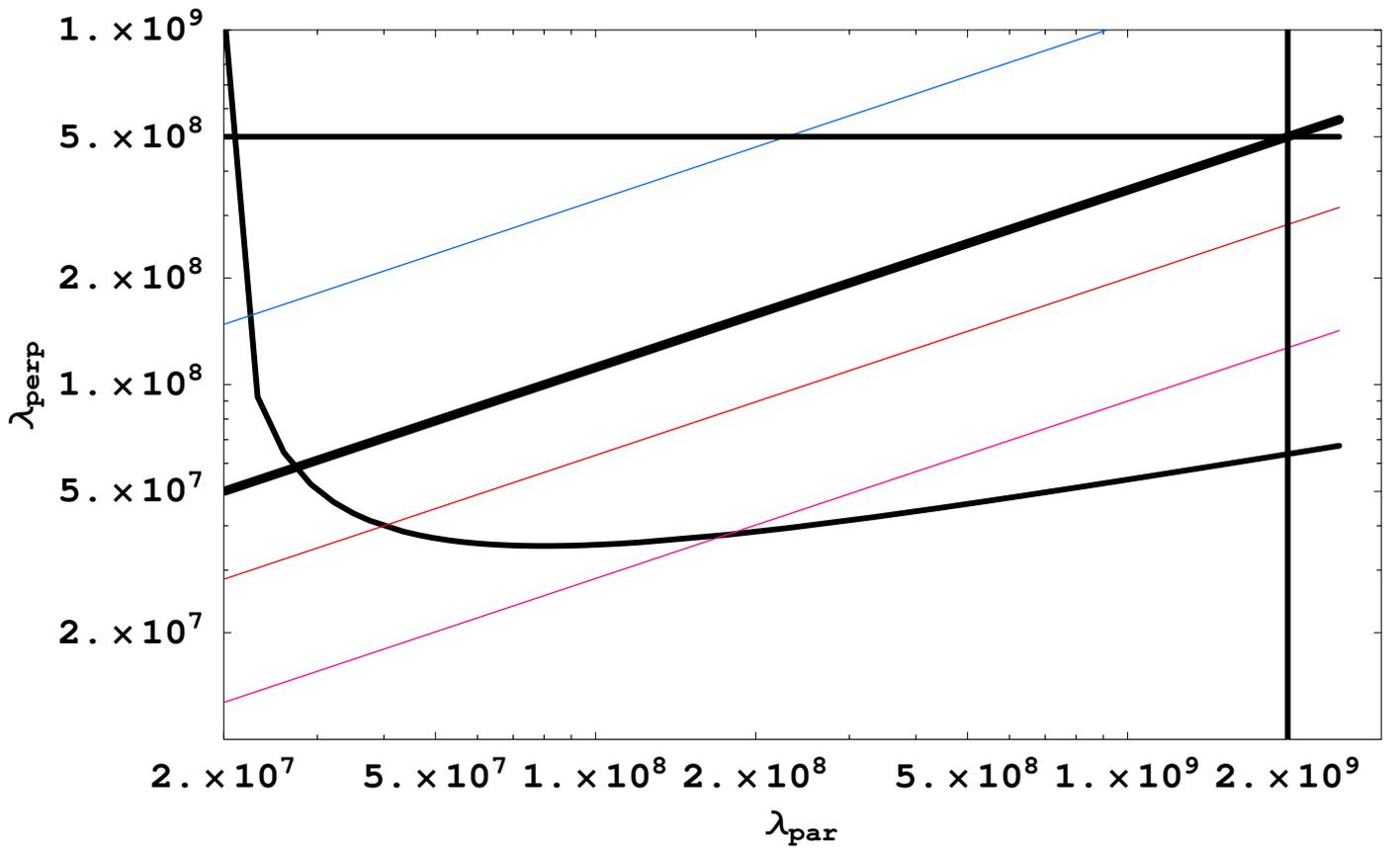}
\caption[]{Kubo number, $K$, in the region of $(\lambda_z,\lambda_\perp)$ space allowed by
observations, bounded by $\lambda_\perp = W_0$ (solid
horizontal line), $\lambda_z = L_0$ (solid vertical line) and $\delta B/B_0 = 1$ (solid curve).
Kubo number $K$ increases downwards in the figure. The thick black line shows $K=0.467$,
and the three coloured lines are drawn for $K = 0.14$, 1 and 4.67.
}
\label{fig:lengths}
\end{figure*}

With $D_m$ known and assumed values of $\lambda_z$ and $\lambda_\perp$, Eq.~(\ref{dminterp}) yields a cubic equation that may
be solved exactly for $(\delta B/B_0)^2$ (although the resulting expression is not particularly informative). Putting $\lambda_z = L_0$
and $\lambda_\perp = W_0$ gives $\delta B/B_0 = 0.12$, the minimum value consistent with $D_m$. In Figure~\ref{fig:lengths}
we show the allowed region of ($\lambda_z,\lambda_\perp$) space, bounded by $\lambda_z = L_0$, $\lambda_\perp = W_0$ and $\delta B/B_0 = 1$.
From Eq.~(\ref{eq:lperp}) in the case $\delta B/B_0 = 1$, we can show that $\lambda_\perp$ has a minimum value of
$4 D_m/3^{3/4} = 1.75D_m = 3.5\times10^7$ cm here.

\begin{figure*}
\centering
\includegraphics[width=\textwidth]{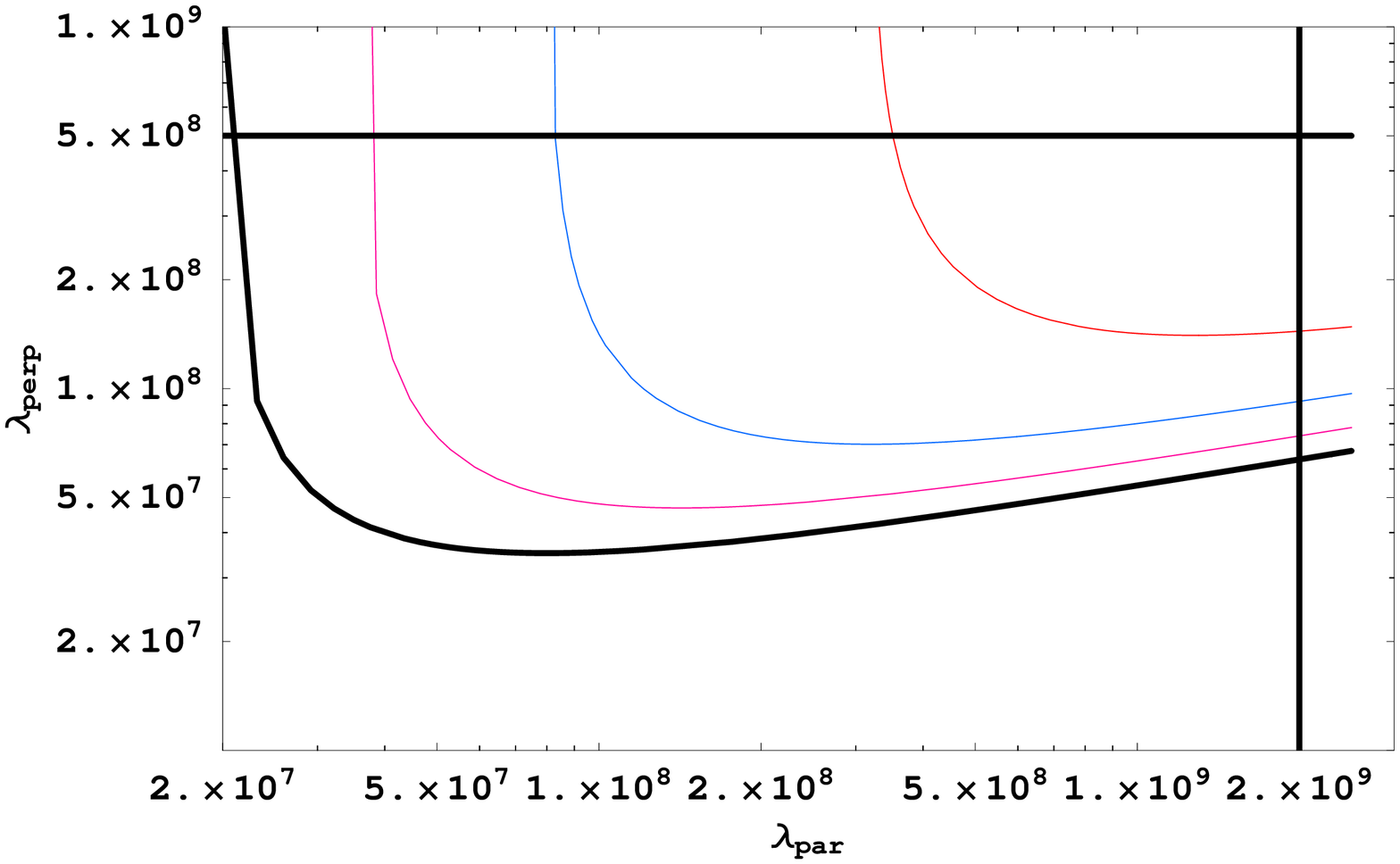}
\caption[]{Contours of  $\delta B/B_0$ in $(\lambda_z,\lambda_\perp)$ space allowed by the observations
$\delta B/B_0= 0.25$ (red),  $\delta B/B_0= 0.5$ (blue) $\delta B/B_0= 0.75$ (magenta).  As in figure \ref{fig:lengths},  $(\lambda_z,\lambda_\perp)$ space
bounded by $\lambda_\perp = W_0$ (solid horizontal line), $\lambda_z = L_0$ (solid vertical line)
and $\delta B/B_0 = 1$ (solid curve) as the limits.
}
\label{fig:lengthsB}
\end{figure*}

The energy in perturbations must be at least $\sim 1\%$ of the energy of the background field.
The possibility remains of a much higher energy contained in the turbulent perturbations (see Figure \ref{fig:lengthsB}),
even sufficient to power the whole flare if $\delta B/B_{0}\simeq 1$. In most of the allowed region $K$ is
of the order of unity $0.1\lesssim K \lesssim10$ (see Figure \ref{fig:lengths}). Nonetheless, we can have significant
anisotropy ($\lambda_z$ and $\lambda_\perp$ substantially different from one another)
without also having $K >>1$. Correlation lengths may not, however, be very much less than the natural
scales of the magnetic loop.

\section{Discussion and summary}

The theory by \citet{1978PhRvL..40...38R}
was applied by \citet{2006ApJ...646..615G} to the thermal loops observed by the TRACE spacecraft
in the extreme ultraviolet range to determine the magnetic turbulence level in thermal loops.
The authors found a level of the order of $\delta B/B_{0}\simeq 0.025-0.075$.
More recently it was pointed out by \citet{2010ApJ...719.1912B}
that by taking into account certain features related to the Kubo number,
the higher value of the order of $\delta B/B_{0}\simeq 0.05-0.7$ could instead be
inferred from the data, values high enough for field-line braiding to lead to sufficient continuous,
small scale reconnection events to account for coronal heating. It should be noted that these estimates for quiet non-flaring
loops appear to be as high as the flaring loop estimates. Furthermore, these
high values of $\delta B/B_{0}\simeq 0.7$ seem to contradict
the observations of non-thermal broadening if interpreted as
turbulent velocities. The corresponding MHD turbulence velocities ${\rm v}\sim B_{\perp}/B_{0} {\rm v}_{A}
\simeq 50-700$~km/s for ${\rm v}_{A}\simeq 1000$~km/s appear much higher
than the typical velocities of tens of km/s inferred from non-thermal line broadening
\cite[][]{1997SoPh..173..243D,1999ApJ...513..969H,2009ApJ...705L.208I}.
For the solar flare conditions discussed by \citet{2011ApJ...730L..22K},
the non-thermal broadening is instead measured in the range of 100-200 km/s \citep{1982SoPh...78..107A,1983SoPh...86...49D,1989ApJ...344..991F,1999A&A...342..279P}
so, $\delta B/B_{0}\simeq 0.1-0.2$ is indeed consistent with these observations.
We note that the thermal particles are more likely to be influenced
by $\mathbf{E}_{\perp}\times \mathbf{B}$ drift, which is dominant
for particles whose speeds is comparable to or less than the Alfven speed.
The observed
appearance of TRACE loops would then constrain $\delta B/B_{0}$ to even lower
values than those found by \citet{2006ApJ...646..615G}.
Alternatively, the ratio of magnetic and kinetic energies in the turbulence is far from
unity and hence the simple relation ${\rm v}\sim B_{\perp}/B_{0} {\rm v}_{A}$
is not applicable.

We considered the cross-field transport of fast
electrons inside coronal loops.
Our analysis was based on a novel method, which exploits the \emph{RHESSI} imaging capabilities
for determining the
value of the magnetic diffusion coefficient in thick target loops.
By ``thick target'' we mean that the flaring loop is dense enough
to guarantee that the electrons remain in the loop while they are accelerated
and emit HXRs, and hence that they are well-observed with X-ray
imaging instruments.
Various possible regimes of cross-field transport of non-thermal electrons
were discussed and applied to the interpretation of the data.
The importance of the Kubo number $K$ as a governing parameter was emphasized
and results applicable to both the quasilinear ($K\ll 1$)
and trapping limits ($K\gg 1$) were collected.

The combination of theory and observation allows us to place interesting constraints
on the relative level of magnetic fluctuations and on the Kubo number in flaring loops. These
are summarized in Figs.(1)-(2). By identifying parallel and perpendicular correlation
lengths with the two integral scales of the visible HXR loop, we found
$\delta B/B_{0}\simeq 0.1$ and also $K\simeq 0.4$. This
quasilinear estimate for  $\delta B/B_{0}$ shows that magnetic fluctuations
with energy of at least $\sim 1\%$ of the energy of the background field can be inferred
from measurements of the magnetic diffusion coefficient.

We note that although the size of the HXR emitting region
is likely to be governed by parallel and perpendicular transport, continuing theoretical
effort is needed to describe the electron dynamics more accurately, in particular regarding the
treatment of the impact of energy loss, acceleration and non-linearity
on transport.

\begin{acknowledgements}

The authors are grateful to G. Fleishman
and V. Nakariakov for valuable discussions.
This work is supported by a STFC rolling grant (NHB, EPK, ALM).
Financial support by the European Commission through the HESPE
Network is gratefully acknowledged.

\end{acknowledgements}

\bibliographystyle{aa}
\bibliography{refs}

\begin{thebibliography}{49}
\expandafter\ifx\csname natexlab\endcsname\relax\def\natexlab#1{#1}\fi

\bibitem[{{Antonucci} {et~al.}(1982){Antonucci}, {Gabriel}, {Acton},
  {Leibacher}, {Culhane}, {Rapley}, {Doyle}, {Machado}, \&
  {Orwig}}]{1982SoPh...78..107A}
{Antonucci}, E., {Gabriel}, A.~H., {Acton}, L.~W., {et~al.} 1982, \solphys, 78,
  107

\bibitem[{{Aschwanden}(2002)}]{2002SSRv..101....1A}
{Aschwanden}, M.~J. 2002, \ssr, 101, 1

\bibitem[{{Aschwanden} {et~al.}(2002){Aschwanden}, {Brown}, \&
  {Kontar}}]{2002SoPh..210..383A}
{Aschwanden}, M.~J., {Brown}, J.~C., \& {Kontar}, E.~P. 2002, \solphys, 210,
  383

\bibitem[{{Balescu}(2000{\natexlab{a}})}]{2000PPCF...42....1B}
{Balescu}, R. 2000{\natexlab{a}}, Plasma Physics and Controlled Fusion, 42, 1

\bibitem[{{Balescu}(2000{\natexlab{b}})}]{2000EJPh...21..279B}
{Balescu}, R. 2000{\natexlab{b}}, European Journal of Physics, 21, 279

\bibitem[{{Bian} {et~al.}(2010){Bian}, {Kontar}, \& {Brown}}]{bian}
{Bian}, N.~H., {Kontar}, E.~P., \& {Brown}, J.~C. 2010, \aap, 519, A114+

\bibitem[{{Bitane} {et~al.}(2010){Bitane}, {Zimbardo}, \&
  {Veltri}}]{2010ApJ...719.1912B}
{Bitane}, R., {Zimbardo}, G., \& {Veltri}, P. 2010, \apj, 719, 1912

\bibitem[{{Brown}(1971)}]{1971SoPh...18..489B}
{Brown}, J.~C. 1971, \solphys, 18, 489

\bibitem[{{Brown} {et~al.}(2002){Brown}, {Aschwanden}, \&
  {Kontar}}]{2002SoPh..210..373B}
{Brown}, J.~C., {Aschwanden}, M.~J., \& {Kontar}, E.~P. 2002, \solphys, 210,
  373

\bibitem[{{Brown} \& {Kontar}(2005)}]{2005AdSpR..35.1675B}
{Brown}, J.~C. \& {Kontar}, E.~P. 2005, Advances in Space Research, 35, 1675

\bibitem[{{Bykov} \& {Fleishman}(2009)}]{2009ApJ...692L..45B}
{Bykov}, A.~M. \& {Fleishman}, G.~D. 2009, \apjl, 692, L45

\bibitem[{{Bykov} \& {Toptygin}(1993)}]{1993PhyU...36.1020B}
{Bykov}, A.~M. \& {Toptygin}, I. 1993, Physics Uspekhi, 36, 1020

\bibitem[{{Chiueh} \& {Zweibel}(1987)}]{1987ApJ...317..900C}
{Chiueh}, T. \& {Zweibel}, E.~G. 1987, \apj, 317, 900

\bibitem[{{Corrsin}(1959)}]{1959AdGeo...6..441C}
{Corrsin}, S. 1959, Advances in Geophysics, 6, 441

\bibitem[{{Doschek}(1983)}]{1983SoPh...86...49D}
{Doschek}, G.~A. 1983, \solphys, 86, 49

\bibitem[{{Doyle} {et~al.}(1997){Doyle}, {O'Shea}, {Erdelyi}, {Dere}, {Socker},
  \& {Keenan}}]{1997SoPh..173..243D}
{Doyle}, J.~G., {O'Shea}, E., {Erdelyi}, R., {et~al.} 1997, \solphys, 173, 243

\bibitem[{{Fludra} {et~al.}(1989){Fludra}, {Bentley}, {Lemen}, {Jakimiec}, \&
  {Sylwester}}]{1989ApJ...344..991F}
{Fludra}, A., {Bentley}, R.~D., {Lemen}, J.~R., {Jakimiec}, J., \& {Sylwester},
  J. 1989, \apj, 344, 991

\bibitem[{{Galloway} {et~al.}(2006){Galloway}, {Helander}, \&
  {MacKinnon}}]{2006ApJ...646..615G}
{Galloway}, R.~K., {Helander}, P., \& {MacKinnon}, A.~L. 2006, \apj, 646, 615

\bibitem[{{Hamilton} \& {Petrosian}(1992)}]{1992ApJ...398..350H}
{Hamilton}, R.~J. \& {Petrosian}, V. 1992, \apj, 398, 350

\bibitem[{{Hara} \& {Ichimoto}(1999)}]{1999ApJ...513..969H}
{Hara}, H. \& {Ichimoto}, K. 1999, \apj, 513, 969

\bibitem[{{Hauff} {et~al.}(2010){Hauff}, {Jenko}, {Shalchi}, \&
  {Schlickeiser}}]{2010ApJ...711..997H}
{Hauff}, T., {Jenko}, F., {Shalchi}, A., \& {Schlickeiser}, R. 2010, \apj, 711,
  997

\bibitem[{{Hurford} {et~al.}(2002){Hurford}, {Schmahl}, {Schwartz}, {Conway},
  {Aschwanden}, {Csillaghy}, {Dennis}, {Johns-Krull}, {Krucker}, {Lin},
  {McTiernan}, {Metcalf}, {Sato}, \& {Smith}}]{2002SoPh..210...61H}
{Hurford}, G.~J., {Schmahl}, E.~J., {Schwartz}, R.~A., {et~al.} 2002, \solphys,
  210, 61

\bibitem[{{Imada} {et~al.}(2009){Imada}, {Hara}, \&
  {Watanabe}}]{2009ApJ...705L.208I}
{Imada}, S., {Hara}, H., \& {Watanabe}, T. 2009, \apjl, 705, L208

\bibitem[{Isichenko(1992)}]{RevModPhys.64.961}
Isichenko, M.~B. 1992, Rev. Mod. Phys., 64, 961

\bibitem[{{Jokipii}(1966)}]{1966ApJ...146..480J}
{Jokipii}, J.~R. 1966, \apj, 146, 480

\bibitem[{{Kadomtsev} \& {Pogutse}(1979)}]{1979ppcf....1..649K}
{Kadomtsev}, B.~B. \& {Pogutse}, O.~P. 1979, in Plasma Physics and Controlled
  Fusion, Vol.~1, Plasma Physics and Controlled Nuclear Fusion Research 1978,
  Volume 1, 649--662

\bibitem[{{Kontar} {et~al.}(2011){Kontar}, {Hannah}, \&
  {Bian}}]{2011ApJ...730L..22K}
{Kontar}, E.~P., {Hannah}, I.~G., \& {Bian}, N.~H. 2011, \apjl, 730, L22+

\bibitem[{{Kontar} {et~al.}(2010){Kontar}, {Hannah}, {Jeffrey}, \&
  {Battaglia}}]{2010ApJ...717..250K}
{Kontar}, E.~P., {Hannah}, I.~G., {Jeffrey}, N.~L.~S., \& {Battaglia}, M. 2010,
  \apj, 717, 250

\bibitem[{{Kontar} {et~al.}(2008){Kontar}, {Hannah}, \&
  {MacKinnon}}]{2008A&A...489L..57K}
{Kontar}, E.~P., {Hannah}, I.~G., \& {MacKinnon}, A.~L. 2008, \aap, 489, L57

\bibitem[{{Kosugi} {et~al.}(1991){Kosugi}, {Masuda}, {Makishima}, {Inda},
  {Murakami}, {Dotani}, {Ogawara}, {Sakao}, {Kai}, \&
  {Nakajima}}]{1991SoPh..136...17K}
{Kosugi}, T., {Masuda}, S., {Makishima}, K., {et~al.} 1991, \solphys, 136, 17

\bibitem[{{Larosa} {et~al.}(1994){Larosa}, {Moore}, \&
  {Shore}}]{1994ApJ...425..856L}
{Larosa}, T.~N., {Moore}, R.~L., \& {Shore}, S.~N. 1994, \apj, 425, 856

\bibitem[{{Lin}(2006)}]{2006SSRv..124..233L}
{Lin}, R.~P. 2006, \ssr, 124, 233

\bibitem[{{Lin} {et~al.}(2002){Lin}, {Dennis}, {Hurford},
  {et~al.}}]{2002SoPh..210....3L}
{Lin}, R.~P., {Dennis}, B.~R., {Hurford}, G.~J., {et~al.} 2002, \solphys, 210,
  3

\bibitem[{{Lin} {et~al.}(2003)}]{2003AdSpR..32.1001L}
{Lin}, R.~P. {et~al.} 2003, Advances in Space Research, 32, 1001

\bibitem[{{Litvinenko}(2006)}]{2006A&A...452.1069L}
{Litvinenko}, Y.~E. 2006, \aap, 452, 1069

\bibitem[{{Melrose}(1994)}]{1994ApJS...90..623M}
{Melrose}, D.~B. 1994, \apjs, 90, 623

\bibitem[{{Miller} \& {Ramaty}(1987)}]{1987SoPh..113..195M}
{Miller}, J.~A. \& {Ramaty}, R. 1987, \solphys, 113, 195

\bibitem[{{Milovanov}(2009)}]{2009PhRvE..79d6403M}
{Milovanov}, A.~V. 2009, \pre, 79, 046403

\bibitem[{{Mrozek}(2006)}]{2006AdSpR..38..962M}
{Mrozek}, T. 2006, Advances in Space Research, 38, 962

\bibitem[{{P{\'e}rez} {et~al.}(1999){P{\'e}rez}, {Doyle}, {Erd{\'e}lyi}, \&
  {Sarro}}]{1999A&A...342..279P}
{P{\'e}rez}, M.~E., {Doyle}, J.~G., {Erd{\'e}lyi}, R., \& {Sarro}, L.~M. 1999,
  \aap, 342, 279

\bibitem[{{Petrosian} \& {Chen}(2010)}]{2010ApJ...712L.131P}
{Petrosian}, V. \& {Chen}, Q. 2010, \apjl, 712, L131

\bibitem[{{Prato} {et~al.}(2009){Prato}, {Emslie}, {Kontar}, {Massone}, \&
  {Piana}}]{2009ApJ...706..917P}
{Prato}, M., {Emslie}, A.~G., {Kontar}, E.~P., {Massone}, A.~M., \& {Piana}, M.
  2009, \apj, 706, 917

\bibitem[{{Rechester} \& {Rosenbluth}(1978)}]{1978PhRvL..40...38R}
{Rechester}, A.~B. \& {Rosenbluth}, M.~N. 1978, Physical Review Letters, 40, 38

\bibitem[{{Saint-Hilaire} {et~al.}(2010){Saint-Hilaire}, {Krucker}, \&
  {Lin}}]{2010ApJ...721.1933S}
{Saint-Hilaire}, P., {Krucker}, S., \& {Lin}, R.~P. 2010, \apj, 721, 1933

\bibitem[{{Shibata}(1999)}]{1999Ap&SS.264..129S}
{Shibata}, K. 1999, \apss, 264, 129

\bibitem[{{Somov} \& {Kosugi}(1997)}]{1997ApJ...485..859S}
{Somov}, B.~V. \& {Kosugi}, T. 1997, \apj, 485, 859

\bibitem[{{Vlad} {et~al.}(1998){Vlad}, {Spineanu}, {Misguich}, \&
  {Balescu}}]{1998PhRvE..58.7359V}
{Vlad}, M., {Spineanu}, F., {Misguich}, J.~H., \& {Balescu}, R. 1998, \pre, 58,
  7359

\bibitem[{{Xu} {et~al.}(2008){Xu}, {Emslie}, \&
  {Hurford}}]{2008ApJ...673..576X}
{Xu}, Y., {Emslie}, A.~G., \& {Hurford}, G.~J. 2008, \apj, 673, 576

\bibitem[{{Zimbardo} {et~al.}(2000){Zimbardo}, {Pommois}, \&
  {Veltri}}]{2000PhyA..280...99Z}
{Zimbardo}, G., {Pommois}, P., \& {Veltri}, P. 2000, Physica A Statistical
  Mechanics and its Applications, 280, 99

\end{thebibliography}

\end{document}